# SUSY-Nonrelativistic Quantum Eigenspectral Energy Analysis for Squared-Type Trigonometric Potentials Through Nikiforov-Uvarov Formalism

Metin Aktaş [(*)]

**Abstract**

Explicit and analytical bound-state solutions of the Schrödinger equation for squared-form trigonometric potentials within the framework of supersymmetric quantum mechanics (SUSYQM) are performed by implementing the Nikiforov-Uvarov (NU) polynomial procedure. The first step requires a certain action to adopt an appropriate ansatz superpotential $W(x)$ for generating the potential pair as $V_{\pm}(x)$. In the second process, inserting each potential for the one-dimensional Schrödinger equation and solving the hypergeometric differential equation with the NU-method gives rise to normalized wave function descriptions and algebraically corresponds to the characteristic SUSY quantum energy eigenspectrum sets. It is remarkable to note that, when examined parametrically, they are of reliable and applicable forms concerning the mathematical treatment of various physical quantum systems prescribed in relativistic or nonrelativistic contexts.

M. Aktaş (metinaktas01@gmail.com)

[(*)] Ankara Yıldırım Beyazıt University, Engineering and Natural Sciences Faculty, Energy Systems Engineering Department, 06010, Ankara, Turkey

# 1. Introduction

The issue of exact and analytical solvability of any quantum mechanical Hamiltonian system offers practical implications for reviewing and improving the various physical structures in the quantum universe. For this reason, it is one of the desirable characteristics because of providing a mathematical solution model compared to systems that are not explicitly solvable. In other words, the existence of a Hamiltonian problem with an exact solution is also a notable subject to familiarize with the usual nature of some quantum mechanisms. As a result, by applying several particular attempts the trial wave function of the configuration for a given potential is estimated and hence the wave behaviour of the physical structure can also be described. In the literature, there has been remarkable interest in solving of the Schrödinger equation (SE) which extends from one to *N*-dimension particularly in the presence of *Coulomb, harmonic oscillator, inverse square, exponential, hyperbolic* and *trigonometric* potential forms as well as their various combinations. One of the distinguishing efforts is to arrive at possible solutions for trigonometric type potentials because of their importance in the applications of molecular physics and quantum chemistry contexts.

The first familiar example available in the literature is related to the exact solutions of Hamiltonian systems with the quadratic form of the cotangent potential [1]. Numerous studies have also been fulfilled in the context. Exact analytical solutions of the SE for a square tangent potential are given [2]. The classical and quantum mechanical treatment of the square tangent potential is presented by applying the operator algebra procedure [3]. In the recent article, general solutions of the SE for trigonometric one are obtained [4]. SUSYQM formalism and the parity-time (PT)-symmetry concept have also been studied for various forms [5 - 9]. In addition, Flügge worked on analytical solutions of the SE for the Pöschl-Teller potential [10]. Furthermore, the exact and approximate solutions of the Schrödinger, Klein-Gordon**,** and Dirac equations from lower to higher dimensions as well as the solutions to the other physical systems were presented for several types of trigonometric potentials in the literature [11 - 35].

The main motivation of this paper is to parametrically derive the supersymmetric forms of some trigonometric style potentials starting from the originals. In addition, the algebraic procedure known as the NU-approach was applied in the review of this research article to obtain acceptable forms of energy eigenvalue and eigenfunction expressions in the nonrelativistic scheme. We will focus on fully solvable potentials via the conceptual framework of SUSYQM, starting with general solvable trigonometric classes such as the squared forms of *tangent, cotangent*, and *trigonemetric Pöschl-Teller - I* cases, respectively. Then, by applying the Nikiforov-Uvarov (NU) polynomial approach [36], the SUSY quantum mechanical energy eigenvalues and their corresponding eigenfunction solutions for the one-dimensional SE are illustrated explicitly. The results are also examined for the complex-valued energy eigenspectra within the context of *PT*-SUSYQM. It is remarkable here to note that the wave function expressions available for these potentials stand for an orthogonal and hypergeometric type polynomial called also as the Jacobi polynomial of the second kind [37]. Generally speaking, they are in applicable forms to review the implications responsible for the stationary and non-stationary states involved in various physical quantum configurations.

The arrangement of this article is presented as follows: The brief formulation of SUSYQM algebra and the polynomial procedure of Nikiforov-Uvarov are introduced in Sections 2 and 3. In Section 4, straightforward calculations are executed to express the energy spectra and corresponding wave functions for the supermates. Moreover, several graphs depicted for both SUSY energy eigenvalues and the ground state wave functions of these potentials are presented in the study of this paper. Furthermore, a Python programming code is implemented to execute the computational procedures. For simplistic calculations, $\alpha = \hbar = m = 1$ is selected. Section 5 deals with the analysis and discussion of the results.

**2. Core Formulation of SUSY Quantum Mechanics**

Although there is no experimental evidence that SUSY is performed in nature so far, the supersymmetric approach provides a more elegant description of nature. Since its emergence in the scientific field, the ideas of SUSY offer new postulates to some branches of physics for understanding *atomic, molecular, nuclear, statistical, condensed matter*, and other quantum systems. Also, the algebra in SUSY is a graded Lie algebra.

SUSYQM has several aspects. One of the important features is the decay of spontaneous SUSY, while the unbroken SUSY state leads to a degeneration between the energy spectra of fermions and bosons. Another important point is the factorization and hierarchy of Hamiltonians. This procedure is applied to generate potential pair in terms of superpotentials. The ground state wave function is also introduced in analytically solvable quantum mechanical problems. For example, this problem can be considered analytically solvable if a potential has the shape invariance property, *e. g.* quantum harmonic oscillator problem.

SUSYQM formalism gives an insight into why certain one-dimensional equations are analytically solvable. Also, it has a challenge on how to solve some problems. For instance, when having no exact solution, then this theoretical formulation offers new powerful approximation approaches including supersymmetry-inspired WKB approximation, large-$N$ expansion, and variational methods. Not only can it be applied to quantum mechanical systems in one dimension but also to certain quantum problems in three dimensions.

Following the algebraic formulation, if the ground state wave function $\Psi_0(x)$ is known, one can determine the partner potential $V_-(x)$, and then factor the Hamiltonian of the system as follows:

$$H_- = -\frac{\hbar^2}{2m}\frac{d^2}{dx^2} + V_-(x)$$

$$\equiv A^\dagger A, \tag{1}$$

where

$$A = \frac{\hbar}{\sqrt{2m}}\frac{d}{dx} + W(x) \quad \text{and} \quad A^\dagger = -\frac{\hbar}{\sqrt{2m}}\frac{d}{dx} + W(x). \tag{2}$$

It leads to identify

$$V_-(x) = W^2(x) - \frac{\hbar}{\sqrt{2m}}W'(x), \tag{3}$$

where $W(x)$ refers to the superpotential of the system in terms of the ground state function, and $W'(x)$ is the first derivative. Secondly, we define the new (second) Hamiltonian

$$H_+ = -\frac{\hbar^2}{2m}\frac{d^2}{dx^2} + V_+(x)$$

$$\equiv A\,A^\dagger. \qquad (4)$$

Thus the new one is equal to

$$V_+(x) = W^2(x) + \frac{\hbar}{\sqrt{2m}}W'(x). \qquad (5)$$

These potentials in equations (3) and (5) are called as *partner potentials*. In this study, the following form of the SE will be solved analytically by adopting an appropriate superpotential to construct them.

$$H_\pm \Psi = E_\pm \Psi. \qquad (6)$$

Hence, it is written explicitly

$$\frac{d^2\Psi}{dx^2} + \frac{2m}{\hbar^2}[E_\pm - V_\pm]\Psi = 0. \qquad (7)$$

Consequently, the iteration process continues up to constructing the *n-th* degree Hamiltonians. This is the procedure well-known as the *hierarchy of Hamiltonians* within the framework of SUSYQM [5].

### 3. A Procedural Approach to the Nikiforov – Uvarov Method

In this technique, a second order differential equation (ODE) is reduced to the hypergeometric form by adopting a suitable coordinate transformation $x = x(z)$

$$\Psi''(z) + \frac{\tilde{\tau}(z)}{\sigma(z)}\Psi'(z) + \frac{\tilde{\sigma}(z)}{\sigma^2(z)}\Psi(z) = 0. \qquad (8)$$

In the equation, while $\sigma(z)$ and $\tilde{\sigma}(z)$ are polynomials with at most second degree, $\tilde{\tau}(z)$ is a polynomial but having at most first degree. By defining the factorization function

$$\Psi(z) = \phi(z)\,y(z), \qquad (9)$$

then one can get the following equation formed as

$$\sigma(z)\, y''(z) + \tau(z)\, y'(z) + \Gamma\, y(z) = 0, \tag{10}$$

where

$$\frac{\pi(z)}{\sigma(z)} = \frac{d}{dz}(ln\phi(z)), \tag{11}$$

and

$$\tau(z) = \tilde{\tau}(z) + 2\pi(z). \tag{12}$$

Besides, $\Gamma$ is defined

$$\Gamma_n + n\,\tau' + \frac{[n(n-1)\sigma'']}{2} = 0 \quad with \quad n = 0, 1, 2, \dots \tag{13}$$

It is noted that the energy eigenvalues can be obtained from the last equation. Now, we first have to determine $\pi(z)$ and $\Gamma$ by defining

$$k = \Gamma - \pi'(z). \tag{14}$$

Solving the quadratic equation for $\pi(z)$ with the use of equation (14) leads to

$$\pi(z) = \left(\frac{\sigma' - \tilde{\tau}}{2}\right) \pm \sqrt{\left(\frac{\sigma' - \tilde{\tau}}{2}\right)^2 - \tilde{\sigma} + k\sigma}\,. \tag{15}$$

In the last equation $\pi(z)$ is a polynomial with the parameter $z$, and prime factors denotes the differentials at first order. The determination of $k$ plays a critical role to arrange $\pi(z)$ in the calculations. After setting the discriminant of the square root to equal zero, then a general quadratic equation for $k$ is obtained.

The second critical step of this procedure is to determine the wave function by following the expressions respectively. The *Rodriguez relation* is given as

$$y_n(z) = \frac{C_n}{\rho(z)} \frac{d^n}{dz^n} [\sigma^n(z)\, \rho(z)], \tag{16}$$

where $C_n$ is normalizable coefficient obtained from the orthogonality relation, and $\rho(z)$ is the weight function satisfying the relation

$$\frac{d}{dz}[\sigma(z)\,\rho(z)] = \tau(z)\,\rho(z). \tag{17}$$

Equation (16) corresponds to the classical orthogonal polynomials that have numerous significant properties, especially satisfy the orthogonality relation

$$\int_a^b y_n(z)\, y_m(z)\, \rho(z)\, dz = \delta_{nm} = \begin{cases} 1, & n = m \\ 0, & n \neq m. \end{cases} \tag{18}$$

**4. Applications**

The sections given above include the fundamental procedures to achieve the SUSY algebraic solutions of the SE with the help of these approaches.

**4.1. The Squared Tangent Potential (STP) Case**

By transforming $x \to \theta$ the potential with $\alpha$-parameter has the form

$$V(\theta) = V_0\, tan^2(\alpha\theta) \equiv V_0\left[\frac{(e^{i\alpha\theta} - e^{-i\alpha\theta})}{i\,(e^{i\alpha\theta} + e^{-i\alpha\theta})}\right]^2, \qquad |\theta| < \frac{\pi}{2} \tag{19}$$

where $V_0 \to \nu(\nu - 1)$ is potential coefficient with the condition $\nu \geq 1$ [1, 2]. SUSYQM approach employs a proposal for its superpotential

$$W(\theta) = A\, tan(\alpha\theta). \tag{20}$$

When following equations (3) and (5), its counterparts are obtained as

$$V_\pm(\theta) = (A^2 \pm \alpha A)\, tan^2(\alpha\theta) \pm \alpha A. \tag{21}$$

For the potential $V_-(\theta)$, equation (7) can be rewritten

$$\frac{d^2\psi}{d\theta^2} + \frac{2m}{\hbar^2}\{E^{(-)} - [(A^2 - \alpha A)\, tan^2(\alpha\theta) - \alpha A]\}\Psi(\theta) = 0, \tag{22}$$

Here, the superscript in $E$ represents the energy spectrum for $V_-(\theta)$. To solve the last equation, one introduces a transformation $z = sin^2(\alpha\theta)$ with $z \in (0, 1)$, hence we have

$$\Psi''(z) + \frac{\left(\frac{1}{2} - z\right)}{z(1-z)}\Psi'(z) + \frac{1}{z^2(1-z)^2}(-\mathcal{E}\, z^2 + \bar{E}\, z)\Psi(z) = 0, \tag{23}$$

with $\mathcal{E} = (\bar{E} + \bar{A})$, $\bar{E} = m\left(E^{(-)} + \alpha A\right)/2\hbar^2\alpha^2$ and $\bar{A} = m\left(A^2 - \alpha A\right)/2\hbar^2\alpha^2$. Comparing equation (23) with equation (8) term-by-term, one gets the polynomials

$$\tilde{\tau}(z) = \left(\tfrac{1}{2} - z\right), \qquad \sigma(z) = z(1-z), \qquad \tilde{\sigma}(z) = (-\mathcal{E}\, z^2 + \bar{E}\, z). \tag{24}$$

Substituting them into equation (15) and rearranging the process will give

$$\pi(z) = \tfrac{1}{4}(1 - 2z) \pm \sqrt{\left[\left(\tilde{\mathcal{E}} - 4k\right) z^2 + \left(\tilde{E} + 4k\right) z + \tfrac{1}{4}\right]}, \tag{25}$$

where $\tilde{\mathcal{E}} = (1 + 4\mathcal{E})$ and $\tilde{E} = (-1 - 4\bar{E})$.

At this point, the determination of $k$ is important step by using the condition given for equation (15). Straightforward calculations conclude to

$$k_{1,\,2} = -\tfrac{1}{8} - \tfrac{\tilde{E}}{4} \pm \tfrac{1}{8}\sqrt{1 + 4(\tilde{\mathcal{E}} + \tilde{E})}. \tag{26}$$

As a result, inserting each value of $k$ into equation (25) yields the results

$$\pi(z) = \tfrac{1}{4}(1 - 2z) \pm \tfrac{1}{4}\begin{cases} \left[\left(1 + \sqrt{1 + 16\bar{A}}\right)z - 1\right], & k = k_1 \\ \left[\left(1 - \sqrt{1 + 16\bar{A}}\right)z - 1\right], & k = k_2. \end{cases} \tag{27}$$

After choosing an appropriate value of $\pi(z)$, equation (12) is computed as

$$\tau(z) = (1 - 2z) + \tfrac{1}{2}\left[\left(1 - \sqrt{1 + 16\bar{A}}\right)z - 1\right]. \tag{28}$$

In addition, if calculating the first and the second derivatives of $\tau(z)$ and $\sigma(z)$ for equation (13) respectively, one would like to have the superpartner energy eigenvalues $E^{(-)}$ for the potential $V_-(\theta)$ afterwards $E^{(+)}$ for the potential $V_+(\theta)$ from equation (22). Therefore, their discrete energy spectra will become

$$E_n^{(\pm)}(\gamma, \delta_1) = \mp \gamma + \frac{\hbar^2 \alpha^2}{m}\left[2n^2 + n\left(1 + \sqrt{1 + 16\bar{A}}\right) + \frac{1}{4}\left(1 + \sqrt{1 + 16\bar{A}}\right)\right]$$

$$= \mp \gamma + \frac{\hbar^2 \alpha^2}{2m}[4n^2 + (4n+1)\delta_1/2]. \tag{29}$$

In the equation, $\delta_1 = \left(1 + \sqrt{1 + 16\bar{\Lambda}}\right)$, $\bar{\Lambda} = m\,(A^2 \pm \alpha A)/2\hbar^2\alpha^2$ and $\gamma = \alpha A$ are defined.

Currently, the action to calculate the corresponding eigenfunctions is taken into consideration. Again by selecting a suitable value of $\pi(z)$ satisfying the condition $\tau'(z) < 0$, they can be determined for equation (9). After computing $\phi(z)$, the *Rodriguez relation* $y_n(z)$, and the *weight function* $\rho(z)$ in equations (11), (16) and (17) respectively

$$\phi(z) = (1-z)^{\delta_1/4}, \quad y_n(z) = C_{\bar{n}}\, G_n(p,q,z), \quad \rho(z) = 1/\left[z(1-z)^{1-\delta_1/2}\right] \tag{30}$$

are expressed. Here, $C_{\bar{n}}$ is the wave function coefficients and $G_n(p,q,z)$ is the second kind Jacobi polynomials with $p = -1 + \delta_1/2$ and $q = 0$ [37], then the total wave function will be

$$\Psi_n(z) = C_{\bar{n}}\,(1-z)^{\delta_1/4}\, G_n(p,q,z), \quad n = 0,1,2,\ldots \tag{31}$$

The straightforward calculations for the coefficients give the result by means of equation (18)

$$C_{\bar{n}} = \left[\frac{(\bar{n}-1)!(\bar{n}-2+\delta_1/2)!}{(2\bar{n}-2+\delta_1/2)!}\right]\sqrt{\frac{\bar{n}\,(\bar{n}-1+\delta_1/2)}{(2\bar{n}-1+\delta_1/2)}}, \quad \bar{n} = 2,3,4,\ldots \tag{32}$$

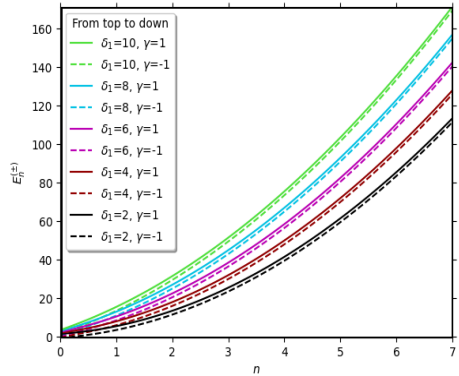

**Fig. 1** SUSY energy eigenvalues versus quantum number *n* for equation (29) with changing $\delta_1$ and $\gamma$.

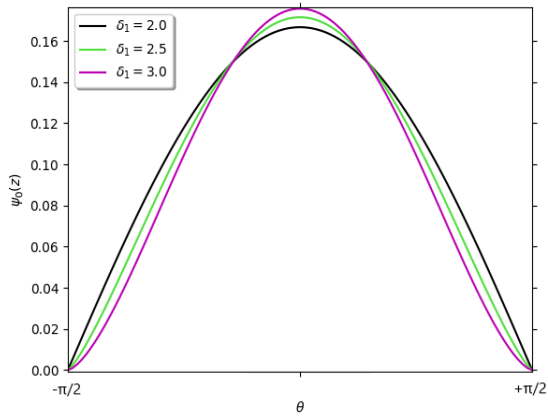

**Fig. 2** Ground state eigenfunctions versus $\theta$ for equation (31) regarding the STP case with changing $\delta_1$.

Figure 1 shows the energy eigenvalue curves with respect to the quantum number *n* varying between 0 and 7 as well as by changing the values of parameters $\delta_1$ and $\gamma$. Solid and dasded lines shown in Figure 1 refer to the the superpartner energy eigenvalues $E^{(-)}$ for the potential $V_-(\theta)$ and $E^{(+)}$ for the potential $V_+(\theta)$ respectively.

Figure 2 shows the plot of the ground state wave functions with respect to $\theta$ by varying the parameter $\delta_1$. They possess a symmetry and maximum value at $\theta = 0$.

### 4.2. The Squared Cotangent Potential (SCP) Case

Now we are going to focus our attention to calculations for the potential defined by the $\alpha$-parameter

$$V(\theta) = V_0 \, cot^2(\alpha\theta) \equiv V_0 \, tan^{-2}(\alpha\theta), \qquad 0 < \theta < \pi \tag{33}$$

where $V_0 = \nu(\nu - 1)$.

With the help of SUSYQM approach the possible form of superpotential will be

$$W(\theta) = A \, cot(\alpha\theta), \tag{34}$$

and its SUSY partners can be generated as

$$V_\pm(\theta) = (A^2 \mp \alpha A) \, cot^2(\alpha\theta) \mp \alpha A. \tag{35}$$

When substituting $V_-(\theta)$ for $E^{(-)}$ into equation (7)

$$\frac{d^2\Psi}{d\theta^2} + \frac{2m}{\hbar^2}\{E^{(-)} - [(A^2 + \alpha A)\cot^2(\alpha\theta) + \alpha A]\}\Psi(\theta) = 0, \tag{36}$$

and applying a transformation $z = \cos^2(\alpha\theta)$ with $z \in (0,1)$, then one gets

$$\Psi''(z) + \frac{\left(\frac{1}{2} - z\right)}{z(1-z)}\Psi'(z) + \frac{1}{z^2(1-z)^2}\left(-\tilde{\mathcal{E}}\, z^2 + \tilde{E}\, z\right)\Psi(z) = 0, \tag{37}$$

where $\tilde{\mathcal{E}} = (\tilde{E} + \tilde{A})$, $\tilde{E} = m\left(E^{(-)} - \alpha A\right)/2\hbar^2\alpha^2$ and $\tilde{A} = m\left(A^2 + \alpha A\right)/2\hbar^2\alpha^2$. This equation has the similar form with equation (23). Therefore, applying the same rigorous algebraic procedure to the equation step-by-step as in the above section will give the SUSY discrete and parametric energy eigenvalues for (35)

$$E_n^{(\pm)}(\gamma, \delta_2) = \pm\gamma + \frac{\hbar^2\alpha^2}{2m}[4n^2 + (4n+1)\delta_2/2]. \tag{38}$$

In the equation, the parameters are $\gamma = \alpha A$, $\delta_2 = \left(1 + \sqrt{1 + 16\tilde{\Lambda}}\right)$ and $\tilde{\Lambda} = m\left(A^2 \mp \alpha A\right)/2\hbar^2\alpha^2$.

Obviously, it is pointed out here that the energy spectra of these partner potentials share similar characteristics as they are in trigonometric forms. However, they differ about a few parameters with opposite signs. This is important and expected result for such periodic potentials (33). In addition to this, careful examination implies that their corresponding eigenfunctions have similar forms but are distinguished by their signs as well as their several parameters.

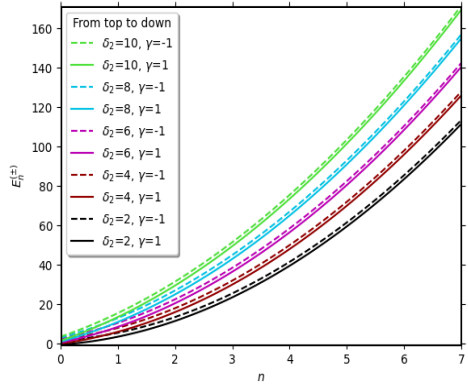

**Fig. 3** SUSY energy eigenvalues versus quantum number *n* for equation (38) with changing $\delta_2$ and $\gamma$.

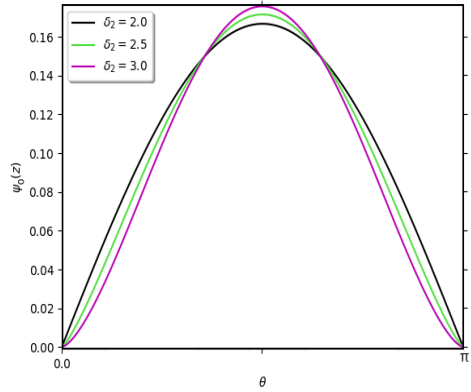

**Fig. 4** Ground state eigenfunctions versus $\theta$ for the SCP case with changing $\delta_2$.

Figure 3 demonstrates the energy eigenvalue curves with respect to the quantum number *n* varying from zero to seven as well as by changing the values of parameters $\delta_2$ and $\gamma$. Dashed and solid lines shown in Figure 3 correspond to the the superpartner energy eigenvalues $E^{(+)}$ for the potential $V_+(\theta)$ and $E^{(-)}$ for the potential $V_-(\theta)$ respectively.

Figure 4 demonstrates the plot of the ground state wave functions with respect to $\theta$ by varying the parameter $\delta_2$. They possess a symmetry and maximum value at $\theta = \frac{\pi}{2}$.

**4.3. Trigonometric Pöschl – Teller Potential (PTP) Case**

In this section, we will cover the calculations for both energy eigenvalues and eigenfunctions. To perform this operation, we first consider the general form of the potential with *α*-parameter as

$$V_{PT}(\theta) = \left[\frac{A}{sin^2(\alpha\theta)} + \frac{B}{cos^2(\alpha\theta)}\right] = [A\, cosec^2(\alpha\theta) + B\, sec^2(\alpha\theta)]$$

$$\equiv \left[ A \left( \frac{e^{i\alpha\theta} - e^{-i\alpha\theta}}{2i} \right)^{-2} + B \left( \frac{e^{i\alpha\theta} + e^{-i\alpha\theta}}{2} \right)^{-2} \right], \tag{39}$$

where the constants are $A = V_0\, \chi(\chi - 1)/2$ and $B = V_0\, \lambda(\lambda - 1)/2$ [10]. While the definition range of $sec(\theta)$ function is between the interval $(-\infty, \infty)$ except for the points $-\frac{\pi}{2}$ and $\frac{\pi}{2}$, that of $cosec(\theta)$ function is again between the same interval except for the points $0$ and $\pi$, that is, $(-\infty, -1] \cup [1, \infty)$.

The ansatz superpotential is introduced

$$W(\theta) = a\, cot(\alpha\theta) + b\, tan(\alpha\theta). \tag{40}$$

Besides its partner potentials provided by the SUSYQM procedure result in the form

$$V_\pm(\theta) = A_1\, cosec^2(\alpha\theta) + B_1\, sec^2(\alpha\theta) - C_1. \tag{41}$$

Here, the coefficients are $A_1 = (a^2 \mp \alpha a)$, $B_1 = (b^2 \pm \alpha b)$ and $C_1 = (a - b)^2$. Inserting the potential $V_-(\theta)$ into the equation (7) and rearranging gives

$$\frac{d^2\Psi}{d\theta^2} + \left[\bar{E} - \bar{a}\, cosec^2(\alpha\theta) - \bar{b}\, sec^2(\alpha\theta)\right]\Psi(\theta) = 0, \tag{42}$$

with $\bar{E} = 2m\left[E^{(-)} + (a - b)^2\right]/\hbar^2$, $\bar{a} = 2m\,(a^2 + \alpha a)/\hbar^2$ and $\bar{b} = 2m\,(b^2 - \alpha b)/\hbar^2$.

Now, we attempt to convert it by introducing a transformation $z = sin^2(\alpha\theta)$ with $z \in (0,1)$ to get

$$\Psi''(z) + \frac{(1-2z)}{2z(1-z)}\Psi'(z) + \frac{1}{4z^2(1-z)^2}\left[-\tilde{E}_1\, z^2 + \tilde{E}_2\, z - \tilde{a}\right]\Psi(z) = 0. \tag{43}$$

The parameters in the equation are $\tilde{E}_2 = (\tilde{E}_1 + \tilde{a} - \tilde{b})$, $\tilde{E}_1 = \bar{E}/\alpha^2$, $\tilde{a} = \bar{a}/\alpha^2$ and $\tilde{b} = \bar{b}/\alpha^2$. If compared with equation (8), the polynomials

$$\sigma(z) = 2z\,(1-z), \qquad \tilde{\tau}(z) = (1-2z), \qquad \tilde{\sigma}(z) = [-\tilde{E}_1\, z^2 + \tilde{E}_2\, z - \tilde{a}], \tag{44}$$

are obtained. By putting them into equation (15), then it can be expressed

$$\pi(z) = \frac{1}{2}(1-2z) \pm \sqrt{\left(\tilde{\mathcal{E}} - 2k\right) z^2 + (\bar{\mathcal{E}} + 2k)\, z + \frac{1}{4}(1 + 4\tilde{a})}, \tag{45}$$

where $\tilde{\mathcal{E}} = 1 - \tilde{E}_1$ and $\bar{\mathcal{E}} = -1 - \tilde{E}_2$. In order to determine $k$ the procedure is put into practice to conclude

$$k_{1,\,2} = \frac{1}{4}\left[(1 + 2\tilde{E}_2) - 2(\tilde{a} + \tilde{b})\right] \pm \frac{1}{4}\sqrt{(1 + 4\tilde{a})(1 + 4\tilde{b})}. \tag{46}$$

When substituting each value of $k$ into equation (45), then one gets

$$\pi(z) = \begin{cases} \frac{1}{2}(1-2z) \pm \frac{1}{2}\left\{\left[\left(\sqrt{1+4\tilde{a}} - \sqrt{1+4\tilde{b}}\right)\right]z - \sqrt{1+4\tilde{a}}\right\}, & k = k_1 \\ \frac{1}{2}(1-2z) \pm \frac{1}{2}\left\{\left[\left(\sqrt{1+4\tilde{a}} + \sqrt{1+4\tilde{b}}\right)\right]z - \sqrt{1+4\tilde{a}}\right\}, & k = k_2. \end{cases} \quad (47)$$

In equation (12) preference of the proper value of $\pi(z)$ yields

$$\tau(z) = 2(1-2z) - \left\{\left[\left(\sqrt{1+4\tilde{a}} + \sqrt{1+4\tilde{b}}\right)\right]z - \sqrt{1+4\tilde{a}}\right\}. \quad (48)$$

Also, after solving equation (13) by using (47) and (48), the common energy eigenvalues are obtained in a parametric manner as well as in a satisfactorily compact form.

$$E_n^\pm(\nu_1, \nu_2) = \frac{\hbar^2\alpha^2}{2m}\{(2n+1)[(2n+1) + (\sqrt{\nu_1} + \sqrt{\nu_2})]$$

$$+ \frac{1}{2}[(1+\sqrt{\nu_1 \nu_2}) + 2(\tilde{a}+\tilde{b})]\} - (a-b)^2, \quad (49)$$

where $\nu_1 = (1+4\tilde{a})$, $\nu_2 = (1+4\tilde{b})$, $\tilde{a} = 2m\,A_1/\hbar^2\alpha^2$ and $\tilde{b} = 2m\,B_1/\hbar^2\alpha^2$.

The next step is necessary to determine the corresponding eigenfunctions. Before starting to calculate we can select a pertinent value of $\pi(z)$ in order for computing $\phi(z)$, $y_n(z)$ and $\rho(z)$ successively by following equations (11), (16) and (17) to be expressed as

$$\phi(z) = [z(1-z)]^{(1+\mu)/2}, \qquad y_n(z) = \bar{C}_n\, G_n(\bar{p}, \bar{q}, z), \qquad \rho(z) = [z(1-z)]^\mu, \quad (50)$$

with $\mu = \left(\sqrt{1+4\tilde{a}} + \sqrt{1+4\tilde{b}}\right)/2$.

Finally, the general form of the wave functions will be available for the trigonometric PTP case

$$\Psi_n(z) = \bar{C}_n\, [z(1-z)]^\mu\, G_n(\bar{p}, \bar{q}, z), \quad n = 0, 1, 2, \dots \quad (51)$$

In this equation, the parameters are $\bar{p} = (1+2\mu)$, $\bar{q} = (1+\mu)$, $z = sin^2(\alpha\theta)$, and $G_n(\bar{p}, \bar{q}, z)$ stands for the Jacobi polynomials of the second kind [37]. The straightforward algebraic process also offers calculating the coefficients of wave function by following equation (18) to get the form

$$\bar{C}_n = \frac{(n+\mu)!}{[(2n+2\mu-1)+1]!}\sqrt{\frac{n!(n+2\mu)!}{(2n+2\mu+1)}}. \tag{52}$$

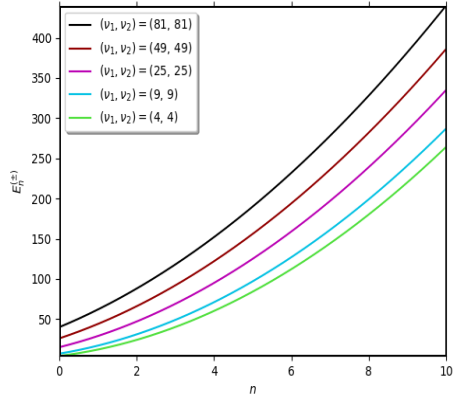

**Fig. 5** SUSY energy eigenvalues versus quantum number *n* for equation (49) with changing $\nu_1$ and $\nu_2$.

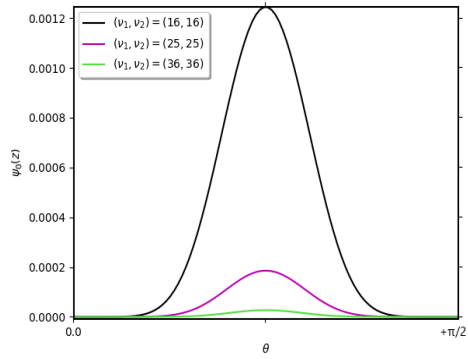

**Fig. 6** Ground state eigenfunctions versus $\theta$ for equation (51) regarding the trigonometric PTP case with changing $\nu_1$ and $\nu_2$.

Figure 5 displays the energy curves with respect to the quantum number *n* changing between 0 and 10 by varying the parameters $\nu_1$ and $\nu_2$. As values of $\nu_1$ and $\nu_2$ increase, energy eigenvalues also increase.

Figure 6 displays the plot of the ground state wavefunctions with respect to $\theta$ by altering $v_1$ and $v_2$. Each function has a symmetry at the midpoint between 0 and $+\pi/2$ and a maximum value.

**Conclusions**

In the study of this article, exact eigensolutions of the Schrödinger equation for SUSY-form trigonometric class potentials such as the squared *tangent, cotangent*, and *Pöschl-Teller - I* cases with $\alpha$-parameter are obtained analytically by applying the Nikiforov-Uvarov (NU) method. Both the wavefunctions and their corresponding energy spectra have explicit forms. They may suggest studying on the behaviour of several quantum mechanical systems. Some of the graphs drawn for both SUSY energy eigenvalues and ground state wave functions are also presented in this paper. In addition, possible outcomes under various conditions are analyzed and reported as follows:

When two cases for the squared trigonometric potentials like STP and SCP are compared in terms of their parametric energy eigenvalues and corresponding eigenfunctions, it is noted that they have similar characteristics. Another remarkable result for them is the existence of a revealing parameter to describe the potential pair if compared to the original ones, as it may serve some noticeable quantum mechanical perspectives.

Only when $\alpha = 0$ is selected, for example, it refers to the free-particle problem both in its original STP form in (19) and its counterparts in (21). However, under the same condition, SCP and its potential pair in equations (33) and (35) gives no possible solution due to infinity.

If $\alpha = 1$ and $\theta = 0$ are chosen, the original form of STP (19) gives a quantum free-particle problem but its one of the potentials in (21) $V_+(\theta) = +A$ refers to the quantum problem for the step potential and the other one $V_-(\theta) = -A$ corresponds to that of the quantum well. On the other hand, under the same condition SCP (33) and its potential pair (35) possess no solution due to infinity. When $\alpha = 1$, at $\theta = \pi/2$ whereas STP and its partners have no solution, SCP and its potential pair has. Also, while the original form of SCP refers to the quantum free-particle problem, one of two counterparts $V_+(\theta = \pi/2) = -A$ refers to the the potential well one, and the other counterpart $V_-(\theta = \pi/2) = +A$ corresponds to the problem of the step potential. Consequently, when their non-relativistic solutions are examined, they yield SUSY quantum eigenspectral configurations. It is also important here to note that, for $\alpha = -1$, equations (21) and (35) transform to the SUSY partners as $V_\pm \rightarrow V_\mp$.

When we analyze the relationships between the general form of the potential (39) as well as its counterpart (41) for the trigonometric PTP case, then we deduce some of the following results:

In the first case, at $\chi = 1$, the second term in (39) reveals whereas at $\lambda = 1$, the first term exists. In the second case, for $\theta = 0, \pi/2, \pi, 3\pi/2$ the potentials in (39) and (41) become singular due to infinity. That is, when at $\theta = 0$ and $\pi$ the first term causes to singularity, whereas at $\theta = \pi/2, 3\pi/2$ the second term gives rise to singularity. Therefore, they are bounded between these values. In addition, for $\alpha = 0$, the same result comes into view.

To illustrate the other point, if $\alpha = 1$, and $\theta \in (0, \pi/2)$ are considered, the significance of the potentials generated by performing the SUSYQM formalism in (41) will exist. In other words, if the coefficients of these potentials (41) are compared with those of the original one in (39), it will ensure some peculiar forms according to the concept of supersymmetry.

In addition, each potential gets new forms by influencing of *PT*-symmetric quantum systems. They offers *PT-symmetry* condition as $\alpha \rightarrow i\alpha$. When applied to the superpartner potentials, they take the new forms with complex-valued $V_\pm(\theta) = [-(A^2 \pm i\alpha A) \tanh^2(\alpha\theta)] \pm i\alpha A$, and $V_\pm(\theta) = [-(A^2 \mp i\alpha A) \coth^2(\alpha\theta)] \mp i\alpha A$ for (STP) and (SCP) cases respectively. Furthermore, under the same circumstances, the trigonometric (PTP) possesses a new form $V_\pm(\theta) = [-(a^2 \pm i\alpha a) \text{cosech}^2(\alpha\theta) + (b^2 \pm i\alpha b) \text{sech}^2(\alpha\theta)] - (a-b)^2$. This assumption implies us the determination of the *PT*-supersymmetric quantum mechanical (*PT*-SUSYQM) energy bound-spectra solutions for these potentials by rearranging the parameters of the eigenvalue results in (29), (38) and (49) as well as the corresponding total eigenfunction results in (31) and (51) respectively.

Finally, it is possible to note that these results may be applicable to the interpretation of quantum behaviour of the physical systems as well as a mathematical model for acceptable solutions to some other quantum mechanical structures.

**Acknowledgment**

It is acknowledged that this study is dedicated to the pretty memory of Tuba Sare Aktaş who will forever live in our hearts.